\documentclass{ws-ijns}
\usepackage[english]{babel}
\usepackage[super,sort,compress]{cite}
\usepackage{xcolor}
\usepackage[verbose,hypertexnames=false]{hyperref}
\usepackage{booktabs}
\usepackage{multirow}
\usepackage{wasysym}

\usepackage{graphicx}
\usepackage{float}
\usepackage{threeparttable}
\usepackage{threeparttablex}
\usepackage{tabularx}
\usepackage{orcidlink}
\usepackage[ruled,vlined,linesnumbered]{algorithm2e}
\usepackage{subcaption}
\usepackage{enumitem}
\usepackage{here}

\hypersetup{colorlinks=false,allbordercolors=blue,pdfborderstyle={/S/U/W 1}}

\makeatletter
\def\wsdraftnote{}  
\makeatother

\begin{document}

\catchline{}{}{}{}{}


\title{HiFIRec Towards High-Frequency yet Low-Intention Behaviors for Multi-Behavior Recommendation}

\author{Ruiqi~Luo, Ran~Jin}
\address{School of Computer Science and Artificial Intelligence, Wuhan Textile University\\Wuhan 430200, China\\
rqluo@.wtu.edu.cn, 2415063008@.wtu.edu.cn\\}
\author{Kaixi~Hu\orcidlink{0000-0002-6774-8510}\thanks{Corresponding author.}}
\address{School of Computer Science and Artificial Intelligence, Wuhan Textile University\\Hubei Key Laboratory of Transportation Internet of Things, Wuhan University of Technology\\Wuhan 430200, China. kxhu@wtu.edu.cn\\}
\author{Xiaohui Tao\orcidlink{0000-0002-0020-077X}}
\address{School of Mathematics, Physics, and Computing, University of Southern Queensland\\Toowoomba 4350, Australia. Xiaohui.Tao@unisq.edu.au\\}
\author{Lin Li\orcidlink{0000-0001-7553-6916}\thanks{Corresponding author.}}
\address{School of Computer Science and Artificial Intelligence, Wuhan University of Technology\\Wuhan 430070, China. cathylilin@whut.edu.cn\\}

\maketitle

\begin{abstract}
Multi-behavior recommendation leverages multiple types of user–item interactions to address data sparsity and cold-start issues, providing personalized services in domains such as healthcare and e-commerce. Most existing methods utilize graph neural networks to model user intention in a unified manner, which inadequately considers the heterogeneity across different behaviors. Especially, high-frequency yet low-intention behaviors may implicitly contain noisy signals, and frequent patterns that are plausible while misleading, thereby hindering the learning of user intentions. To this end, this paper proposes a novel multi-behavior recommendation method, HiFIRec, that corrects the effect of high-frequency yet low-intention behaviors by differential behavior modeling. To revise the noisy signals, we hierarchically suppress it across layers by extracting neighborhood information through layer-wise neighborhood aggregation and further capturing user intentions through adaptive cross-layer feature fusion. To correct plausible frequent patterns, we propose an intensity-aware non-sampling strategy that dynamically adjusts the weights of negative samples. Extensive experiments on two benchmarks show that HiFIRec relatively improves HR@10 by 4.21\%–6.81\% over several state-of-the-art methods.

\end{abstract}


\vspace*{-2pc}

\begin{multicols}{2}

\section{Introduction}
\label{s:introduction}
The development of e-infrastructures (e.g., mobiles, clouds) enables the recording of diverse human behaviors beyond the constraints of time and space. Recommendation systems are an effective tool that delivers tailored services by analyzing historical behaviors. For example, healthcare recommendation~\cite{liang2025enhancing} provides personalized interventions from patients' medical records while ensuring privacy and interpretability. Moreover, other open-world web recommendation systems, including those in e-commerce~\cite{DBLP:journals/eswa/JyothiLN25} and social media~\cite{DBLP:conf/dasfaa/ChenFGLWLZ23}, leverage abundant log data to enhance personalized user experience. Traditional recommendation methods~\cite{zhou2023contrastive,qi2023traditional} usually suffer from data sparsity and cold start problems, since they rely on a single type of behavior. By fully exploiting distinct user-item interaction behaviors (e.g., view, add, and purchase), multi-behavior recommendation methods~\cite{DBLP:conf/recsys/ElsayedRS24,DBLP:journals/tois/YanCGSLSL23} are able to capture accurate user intentions, thereby delivering effective services.

\begin{figure*}[t]
 	\centering
 	\includegraphics[width=0.98\linewidth]{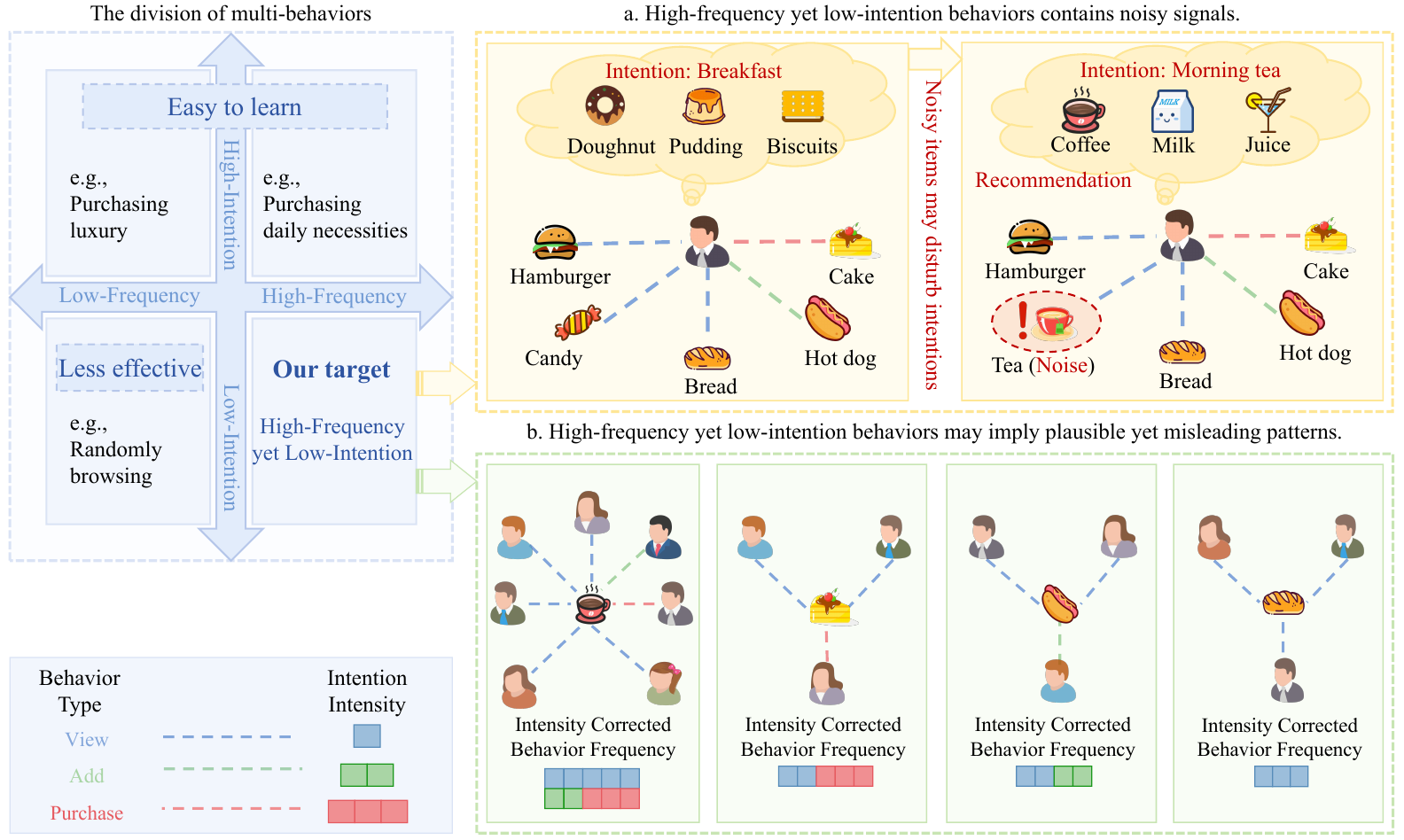}
	\caption{An illustration of high-frequency yet low-intention behaviors. High-frequency yet low-intention behaviors are the primary sources of noisy signals and plausible frequent patterns, to which the unified modeling of multi-behaviors is particularly prone. The Sub-figure (a) shows the noisy item ``tea'' may disturb intentions. Sub-figure (b) indicates the behaviors with different intention intensity may distort frequency-based distribution.}
	\label{fig:fig1}
\end{figure*}

Modeling the heterogeneity of interaction behaviors between users and items is the fundamental challenge in multi-behavior recommendation, since different behavior types inherently carry distinctive user intentions. Early works~\cite{DBLP:conf/ijcai/DingY0QLCJY18,DBLP:conf/icde/Gao0GCFLCJ19,chen2020efficient} mainly follow the paradigm of collaborative filtering (CF), which lacks an explicit encoding of heterogeneous interactions. With the development of deep learning, existing studies can be categorized into sequence-based methods~\cite{DBLP:conf/recsys/ElsayedRS24,DBLP:conf/icde/WuCYM24,DBLP:conf/www/HanWWWLGLC24,DBLP:conf/sigir/YuanG0GLT22} and graph-based methods~\cite{DBLP:journals/ijhcs/LouZJH25,chen2021graph,xu2023multi,lan2024contrastive}. The former mainly focuses on temporal dynamics in user behaviors. For example, Yuan et al.~\cite{DBLP:conf/sigir/YuanG0GLT22} integrate a generator into Transformer~\cite{DBLP:conf/nips/VaswaniSPUJGKP17}, encoding diverse sequential patterns, which is particularly useful in scenarios where temporal dynamics strongly impact outcomes.  Differently, graph-based methods aim to model the high-hop connectivity~\cite{chen2021graph} and the intricate structural property of user-item interaction behaviors. Both groups represent equally promising directions in multi-behavior recommendation. 

Recently, the development of graph representation learning has drawn great attention from the research community. Graph-based multi-behavior recommendation represent users and items as nodes and their interactions as edges. Prior works such as GHCF~\cite{chen2021graph}, utilize graph heterogeneous collaborative filtering to capture high-hop connectivity. However, such methods adopt a unified manner to model intentions underlying types of behavior, improperly assuming their contributions equally. Several works perform differential behavior modeling to model semantic differences between behaviors, including node self-discrimination~\cite{xu2023multi} and contrastive learning~\cite{lan2024contrastive}.

Despite the effectiveness of the above methods, they still do not fully consider the heterogeneity of behavior frequency and underlying intention intensity. As illustrated in Fig.~\ref{fig:fig1}, each behavior can be described along two dimensions - behavior frequency and intention intensity. High-intention behaviors are clearly goal-driven and strongly motivated, making them easier to learn. The effect of low-frequency yet low-intention behaviors may be covered by volume data in statistics. Importantly, \textbf{High-Frequency yet Low-Intention} behaviors account for a large proportion of training data, which requires careful handling. Their effects are summarized, as follows:

\begin{itemize}[left=0pt]
    \item \textbf{Low-intention behaviors potentially contain noisy signals.} In Fig.~\ref{fig:fig1}a, a user may have an inherent interest in purchasing breakfast food. If a noisy item like ``tea'' is inadvertently viewed, it may make models mislead in the intention of morning tea that is often around mid-morning to help refresh before lunch. Therefore, how to effectively correct the effect of noisy signals in  high-frequency yet low-intention behavior is challenging, making it crucial to refine behavior representations. 
    
    \item \textbf{Some frequent behavior patterns may appear plausible yet be misleading.} In Fig.~\ref{fig:fig1}b, the interaction frequency of low-intention behaviors (predominantly ``view'') is higher than that of other behaviors. Overemphasizing such behaviors will mislead models toward frequent patterns that appear plausible yet reflect false intentions. Positive samples (items with which users interact) and negative samples reflect two sides of a behavioral pattern. It is potential to reduce this effect by mining user intentions from negative samples. However, existing non-sampling methods~\cite{10.1145/3373807,chen2020efficient} treat all negative samples equally, disregarding differences in intention intensity.    
\end{itemize}

To this end, we propose a novel multi-behavior recommendation method, termed as \textbf{HiFIRec}, which differentially models behavioral intentions from two perspectives. For the network structure, we propose a hierarchical noise correction module. This module performs layer-wise neighborhood aggregation on user, item, and behavior features. In addition, an adaptive cross-layer feature fusion is introduced to integrate behavioral semantics across different levels, effectively suppressing the noisy signals caused by high-frequency yet low-intention behaviors. For the learning strategy, we introduce an intensity-aware non-sampling strategy to mitigate the effect of plausible frequent patterns. Negative samples are assigned differentiated weights based on the intention intensity associated with different behavior types, thereby reducing the dominant effect of high frequency items during training. 

Our main contributions are as follows:
\begin{itemize}[left=0pt]
\item This pioneer work highlights high-frequency yet low-intention behaviors that potentially contains noisy signals and plausible yet misleading frequent behavior patterns.
\item This paper presents the HiFIRec method, which effectively addresses the effect of high-frequency yet low-intention behaviors through differential representation learning of behavioral intentions.
\item Extensive experiments conducted on Taobao and Beibei datasets show that our proposed HiFIRec achieves superiority over several state-of-the-art baselines with 4.21\%-6.81\%  improvements in terms of HR@10.
\end{itemize}

\section{Related Work}
\label{s:related work}
\subsection{Multi-Behavior Recommendation}

Multi-behavior recommendation leverages different user interaction data (e.g., view, add, and purchase) to capture user intentions. Early recommendation systems mainly relied on content-based~\cite{mahbod2025evaluating,wang2018deep} and collaborative filtering~\cite{zhang2018improved} methods. With the development of deep learning~\cite{lara2021experimental}, recent research~\cite{palumbo2023algorithm} has increasingly focused on two main paradigms. Sequence-based methods~\cite{wu2024personalized} emphasize the temporal dynamics of user behaviors, while graph-based methods emphasize the complex structural relationships between users and items. Since the interests underlying high-frequency yet low-intention behaviors are broadly distributed and prone to interest fluctuation, exhibiting weak temporal dependencies, it is suitable to model their high-hop connectivity through a graph (e.g., GCN~\cite{mao2021ultragcn}, GAT~\cite{tao2020mgat}).

Graph-based multi-behavior recommendation can be divided into two groups. \textbf{The first group} employs unified behavior modeling, uniformly processing different types of behaviors~\cite{mo2025fgcm}. It decomposes the heterogeneous multi-behavior graph into homogeneous sub-graphs through meta-paths or edge partitioning of specific relationships~\cite{wang2019heterogeneous,Zhou2020AWG}. And extends convolutional neural networks to incorporate different behavior information during the message-passing process~\cite{chen2020efficient, chen2021graph}. However, this will be affected by multi-behavioral differential noisy signals and lead to the loss of detailed features of specific behavior patterns. To clearly distinguish differences among multiple behaviors, \textbf{the second group} adopts differentiated behavior modeling~\cite{yang2021hyper,ren2024sslrec,zhang2024temporal,ijcai2022p285,xu2023multi,lan2024contrastive,li2022co,wu2022multi,xuan2023knowledge}. Some works employ self-supervised learning~\cite{ren2024sslrec,xu2023multi,ijcai2022p285}, such as node self-discrimination~\cite{xu2023multi} and self-supervised graph collaborative filtering~\cite{ijcai2022p285}. Some works leverage contrastive learning~\cite{wu2022multi,li2022co,zhang2024temporal}, including contrastive clustering learning~\cite{lan2024contrastive} , non-sampling co-contrastive learning~\cite{li2022co} and hyper meta-path~\cite{yang2021hyper}. More recently, MBRCC~\cite{lan2024contrastive} uses contrastive learning to analyze the impact of auxiliary behaviors on target behaviors, and uses clustering contrast strategies to capture the similarities between users and items. 

The above methods primarily focus on addressing data sparsity~\cite{xu2023multi,lan2024contrastive}, exploring commonalities between target and auxiliary behaviors~\cite{xu2023multi,ijcai2022p285,lan2024contrastive}, and capturing the complex dependencies among multiple behaviors~\cite{yang2021hyper}. Differently, our work uniquely focuses on high-frequency yet low-intention behaviors and simultaneously models the noisy signals and plausible frequent patterns introduced by them.
 
\subsection{Non-sampling Learning}

Non-sampling learning regards all unlabeled samples as negative samples~\cite{10.1145/2911451.2911489, 4781121, 10.1145/2872427.2883090}. It has drawn attention due to its advantages in addressing the data sparsity problem in recommendation tasks. By leveraging the entire training dataset, non-sampling learning can more effectively capture collaborative filtering relationships between users and items, as well as semantic relationships among users, thus improving performance. However, these methods such as traditional non-sampling learning suffer from high time complexity and low efficiency~\cite{10.1145/2911451.2911489}. To improve the efficiency, various methods have been developed, including Alternating Least Squares (ALS)~\cite{10.1145/2911451.2911489, 4781121} and mini-batch Stochastic Gradient Descent (SGD)~\cite{10.1145/3331184.3331192, 10.1145/3373807, xin-etal-2018-batch}. These methods have been widely applied in traditional recommendation models~\cite{10.1145/2911451.2911489,4781121, 10.1145/2872427.2883090}.

ENMF~\cite{10.1145/3373807} is one of the non-sampling strategies studied in neural network-based recommendation systems. Through rigorous mathematical analysis, they significantly reduced the time complexity of learning on the entire dataset at scale. Their follow-up work further verified the superior performance of efficient non-sampling learning~\cite{10.1145/3397271.3401040,liang2023graph}.

While traditional non-sampling methods can alleviate data sparsity, they suffer from high time complexity and low efficiency. Methods like ENMF, which focus solely on item frequency, struggle to account for the impact of high-frequency yet low-intention behaviors. Moreover, improper weighting of negative samples may even exacerbate plausible frequent patterns. Based on non-sampling learning, HiFIRec mitigates the impact of plausible frequent patterns by quantifying the differences in intention intensity among different behaviors and dynamically adjusting the negative samples weights.

\section{Proposed Method}
\label{s:proposed method}

Let $\mathcal{U}=\{u_1,u_2,\dots,u_M\}$ and $\mathcal{V}=\{v_1,v_2,\dots,v_N\}$ denote the sets of users and items with cardinalities $M$ and $N$, respectively. We consider $K$ types of interaction behaviors (e.g., view, add, and purchase), represented as a third-order interaction tensor $\mathcal{I} \in \mathbb{R}^{M\times N\times K}$. For each type of behavior $k \in \{1,...,K\}$, the interaction matrix $I^{(k)} \in \{0,1\}^{M\times N}$ indicates whether a specific user-item interaction of type $k$ occurs. The objective is to predict the probability $y_{u,v}$ of a user $u$ performing a target behavior (e.g., purchase) on the item $v$ , by learning embeddings of user and item that effectively alleviate the noisy signals and plausible frequent patterns introduced by high-frequency yet low-intention behaviors.

\begin{figure*}
 	\centering
 	\includegraphics[width = 1\linewidth]{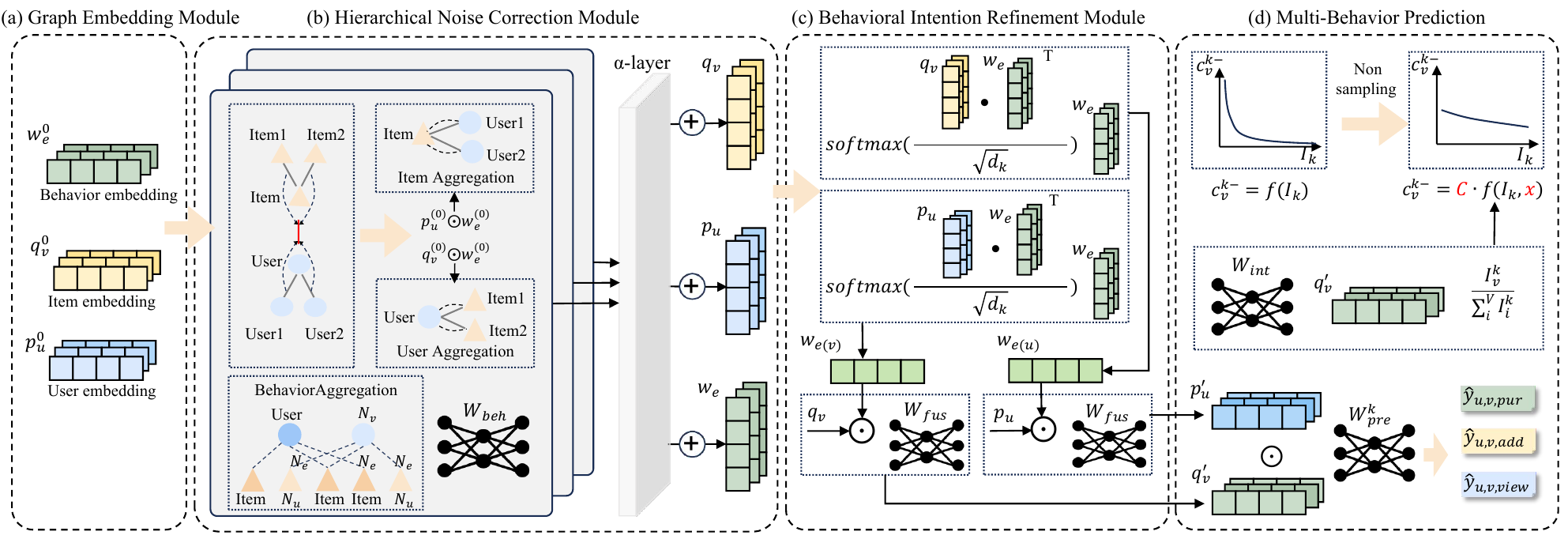}
	\caption{The Overall Framework of HiFIRec. The framework primarily comprises four modules. (a) The Graph Embedding Module provides the representations for users, items, and behaviors. (b) The Hierarchical Noise Correction Module reduces noisy signals through Layer-Wise Neighborhood Aggregation and Adaptive Cross-Layer Feature Fusion. (c) The Behavioral Intention Refinement Module employs an adaptive behavior weighting to mitigate noisy signals effects. (d) The Multi-Behavior Prediction employs a non-sampling strategy to mitigate plausible frequent patterns.}
	\label{fig:fig2}
\end{figure*}

\subsection{Overview}
To cope with high-frequency yet low-intention behaviors, we propose HiFIRec that reduces their effect by hierarchical noise correction and differentiated weight allocation. As shown in shown in Fig.~\ref{fig:fig2}, we first construct user intention representations by jointly modeling users, items, and behaviors, and performs hierarchical noise correction through neighborhood aggregation and cross-layer feature fusion. Furthermore, assigning appropriate weights to different behaviors helps highlight key behavior signals, while adjusting for item frequency reduces interference in behavior modeling.

\subsection{Graph Embedding Module}
To model the heterogeneous nature of the user-item-behavior interaction graph, we first initialize embeddings for users, items, and behavior-specific edges. A user $u \in \mathcal{U}$ and an item $v \in \mathcal{V}$ are mapped into a shared low-dimensional space via learnable matrices $P \in \mathbb{R}^{|\mathcal{U}| \times d}$ and $Q \in \mathbb{R}^{|\mathcal{V}| \times d}$, respectively. To distinguish between different types of user behaviors, we define separate embedding tables for each edge type: $W_{\text{view}}$, $W_{\text{add}}$, and $W_{\text{purchase}}$. This embedding scheme unifies heterogeneous nodes and relations within a consistent vector space, enabling the model to effectively capture diverse interaction semantics during the graph propagation process.

At the initial stage, all embeddings are randomly initialized. Given a user $u$, an item $v$, and a behavior interaction $e$ of type $k$, we transform them into learnable embeddings to construct the heterogeneous behavior graph. The corresponding equations are as follows:
\begin{equation}
\begin{aligned}
p_u^{(0)} &= \text{embedding}(u) = P[u], \\
q_v^{(0)} &= \text{embedding}(v) = Q[v], \\
w_e^{(0)} &=
    \begin{cases}
    W_{\text{view}}[e], & \text{if } e \in \mathcal{E}_{\text{view}}, \\
    W_{\text{add}}[e], & \text{if } e \in \mathcal{E}_{\text{add}}, \\
    W_{\text{purchase}}[e], & \text{if } e \in \mathcal{E}_{\text{purchase}}.
    \end{cases}
\end{aligned}
\end{equation}
where $p_u^{(0)} \in \mathbb{R}^d$ denotes the initial embedding of user $u$, $q_v^{(0)} \in \mathbb{R}^d$ denotes the initial embedding of item $v$, and $w_e^{(0)} \in \mathbb{R}^d$ represents the initial embedding of behavior edge $e$.
 
\subsection{\protect\mbox{Hierarchical Noise Correction Module}}
The noisy signals in high-frequency yet low-intention behaviors may have varying effects on the learning of user intentions at different layers. Hence, the Hierarchical Noise Correction Module aims to establish a differential representation learning process that reduces the effect of noisy signals. It mainly consists of Layer-Wise Neighborhood Aggregation and Adaptive Cross-Layer Feature Fusion.

\subsubsection{Layer-Wise Neighborhood Aggregation}

For the graph representation learning of user intentions in each layer, noisy behaviors may convey misleading or even contradictory information compared to their local neighborhoods. By incorporating contextual neighborhood information, the relevance among different users, items and behaviors can be leveraged to amplify consistent intentions and suppress the interference from noisy behaviors. This enables the model to smooth out the effect of noisy behaviors. Therefore, we aggregate layer-wise neighborhoods for users, items, and behaviors.

As shown in the left part of hierarchical noise correction module in Fig.~\ref{fig:fig2}b, the Aggregation Convolutional Kernel (ACG) is used to update the representations of users and items. Specifically, in each propagation layer, item representations and behavioral representations (e.g., view, add, purchase) are multiplied element-wise to construct user representations incorporate behavioral semantics. Similarly, user representations and behavioral (edge) representations are multiplied element-wise to construct item representations integrate user behavior information. The process is illustrated in the following formulas:
\begin{equation}
\begin{aligned}
 p_u^{(l+1)} = ACG(q_v^{(l)} \odot w_e^{(l)}\quad v, e \in \mathcal{N}_u),\\
 q_v^{(l+1)} = ACG(p_u^{(l)} \odot w_e^{(l)}\quad u, e \in \mathcal{N}_v),
\end{aligned}
\end{equation}
where $\mathrm{ACG}(\cdot)$ represents the aggregation function of the graph convolutional kernel, $l$ denotes the propagation layer, $\mathcal{N}_{u}$ refers to the set of all item nodes connected to user $u$, $\mathcal{N}_{v}$ represents the set of all user nodes connected to item $v$.

In addition to the update of the nodes representation, the representations of the edges are dynamically adjusted in each layer of propagation. We consider the neighborhood edge information of each edge ( other edges associated with the same user or item ). In the $l+1$-th layer, the representation of an edge is updated by aggregating the representations of its adjacent edges. The definition is as follows:
\begin{equation}
	w_e^{(l+1)} = \sigma \left(\sum_{e_{(uv)} \in \mathcal{N}_e} \frac{1}{\left|\mathcal{N}_u + \mathcal{N}_v\right|} W_\mathrm{beh}^{(l)} w_{e_{(uv)}}^{(l)}\right),
\end{equation}%
where $\mathcal{N}_{e}$ represents the set of adjacent edges of edge e, $W_\mathrm{beh}$ is the learnable weight matrix associated with the behavior type,  and $\sigma(\cdot)$ represents the nonlinear activation function applied element-wise.

\subsubsection{Adaptive Cross-Layer Feature Fusion}

Different layers inherently encode different levels of intention features, which play distinct roles in understanding user intentions and suffer from different effects of noisy signals. Layer-wise neighborhood aggregation is insufficient to fully reduce the effect of noisy signals. In particular, noisy signals present in a node or edge can propagate to deeper layers, and simply stacking layers may accumulate these negative effects. In this work, we adopt cross-layer feature fusion, which enables the model to combine high-order intention features from deeper layers with low-order intention features. To further suppress noisy signals, we introduce an adaptive $\alpha$-layer allows the model to learn which layers contain cleaner and more informative intention signals.

As shown in the right part of hierarchical noise correction module in Fig.~\ref{fig:fig2}b, we incorporate the neighbor information to obtain updated user, item, and behavior features at each layer. The features of each layer are multiplied by the corresponding attention weight $\alpha$. By accumulating the weighted features from zero-th layer to $l$-th layer, we derive the fused representations of users, items, and behaviors. This process is illustrated in the following formulas:
\begin{equation}
\begin{aligned}
	w_e = \sum_{l=0}^L \alpha^{(l)} w_e^{(l)}, \\
	p_u = \sum_{l=0}^L \alpha^{(l)} p_u^{(l)}, \\
	q_v = \sum_{l=0}^L \alpha^{(l)} q_v^{(l)}.
\end{aligned}
\end{equation}

Let $w_e^{(l)}$, $p_u^{(l)}$, and $q_v^{(l)}$ denote the representations of edges, users, and items obtained after the $l$-th layer of propagation. The attention vector $\boldsymbol{\alpha} = [\alpha^{(0)}, \alpha^{(1)}, \alpha^{(2)}, \dots, \alpha^{(l)}]$ is trainable and calculated using a softmax function to ensure non-negativity and interpretability: $\alpha^{(l)} = \frac{\exp(\theta^{(l)})}{\sum_{j=0}^{L} \exp(\theta^{(j)})}$,where $\theta^{(l)}$ is a learnable scalar parameter for the $l$-th layer. 

This fusion allows for deeper integration of behaviors and node information, thereby enhancing the model ability to capture complex relationships and patterns within the data. Finally, $p_u$, $q_v$, and $w_e$ the representations of the fused users, items, and behaviors across different layers, are subsequently passed to the following processing module for learning.

\subsection{\protect\mbox{Behavioral Intention Refinement Module}}
The Hierarchical Noise Correction Module is able to correct most of the noisy signals, but there may still be high-frequency yet low-intention behaviors that are not fully suppressed. This is because the module simply fuses the representations of all behavior types without distinguishing their relative importance. To address this, we propose the Behavioral Intention Refinement Module. This module employs a behavioral attention mechanism to emphasize key behaviors in modeling user intentions. To further enhance the representation of user intentions, it integrates various edge information to produce a refined representation.

The Behavioral Intention Refinement Module is shown in Fig.~\ref{fig:fig2}c. We design an adaptive behavior weighting method. This method leverages an attention mechanism~\cite{DBLP:conf/nips/VaswaniSPUJGKP17} to perform a weighted sum over the neighboring edges of users or items. It learns which behavioral edges are more relevant to the users or items in order to construct context-aware edge representations. The resulting weighted behavioral features, $w_{e(u)}$ and $w_{e(v)}$, are then used as inputs for behavior modeling. This process is illustrated in the following formulas:
\begin{equation}
\begin{aligned}
	w_{e(u)} = \sum_{e \in \mathcal{N}_u}{softmax(\frac{p_u \cdot w_e^T}{\sqrt{d_k}})w_e} ,\\
	w_{e(v)} = \sum_{e \in \mathcal{N}_v}{softmax(\frac{q_v \cdot w_e^T}{\sqrt{d_k}})w_e},
\end{aligned}
\end{equation}
where $w_{e(u)}$ and $w_{e(v)}$ represent the corresponding high-order behavioral features of user $u$ and item $v$, respectively, capturing their complex interaction patterns across multiple layers.

We perform integration of user intention features and behavioral contextual information. Specifically, 
the neighbor-aggregated representation $w_{e(u)}$ (obtained in the previous step) is combined with the representation of the user $p_u$ through element-wise multiplication to model interactions. The fused features are then mapped into a more refined representation through a linear transformation. By combining the user historical behavior data, we generate a refined user representation. The calculation process for the refined item representation is the same as that for the user. The relevant formulas are as follows:
\begin{equation}
\begin{aligned}
 {p'}_u = \sigma\left(W_{fus}(p_u \odot w_{e(u)})\right),\\
 {q'}_v = \sigma\left(W_{fus}(q_v \odot w_{e(v)})\right),
 \end{aligned}
\end{equation}%
where $W_{fus}$ is a learnable weight matrix of a linear transformation, ${p'}_u$ is the refined user representation, ${q'}_v$ is the refined item representation.

\subsection{Multi-Behavior Prediction} 
In Fig.~\ref{fig:fig2}d, the multi-behavior prediction module consists of the Intensity-Aware Negative Sampling Strategy and the Joint Multi-Behavior Optimization. 

\subsubsection{\protect\mbox{Intensity-Aware Non-Sampling Strategy}}
To avoid models learning frequent behavior patterns that look plausible yet misleading information, we further mine user intentions from negative samples. However, existing non-sampling methods~\cite{10.1145/2911451.2911489} usually use uniform fixed weights to process all negative samples, which ignores the differences in intention intensity reflected by different behaviors. In this work, we propose an Intensity-Aware Non-Sampling Strategy. The core idea is to dynamically assign differentiated weights to negative samples based on the intention intensity of each behavior type.

ENMF~\cite{10.1145/3373807} employs a non-sampling strategy, assigning fixed weights to positive samples and popularity-based weights to negative ones. While effective at reducing sampling bias, it overlooks the heterogeneity across different behavior types. To balance the differences between different behaviors during the training process, we propose the Intensity-Aware Non-Sampling Strategy.

Formally, for each behavior type $k$, the ENMF loss function is defined as:
\begin{equation}
	\mathcal{L}_k(\Theta) = \mathrm{ENMF} \left(W_{pre}^{k}, {p'}_u, {q'}_v, c_v^{k}\right),
\end{equation}
where $c_v^{k}$ denotes the sample weight. In ENMF, $c_v^{k}$ is divided into positive and negative cases: observed interactions $c_v^{k+}$ are assigned a fixed weight, while unobserved interactions $c_v^{k-}$ are determined directly by item frequency:
\begin{equation}
 c_v^{k} =
 	\begin{cases}
 c_v^{k+}, & \text{if observed}, \\
 c_v^{k-}, & \text{if unobserved}.
 	\end{cases}
\end{equation}

However, the popularity-based approach is not well-suited for multi-behavior data. To address this, we first normalize frequency to align the scales across behaviors:
\begin{equation} 
 c_v^{k-} = C \cdot \frac{ {(f_{v}^{k\text{n}})}^x }{ \sum_{i}^\mathcal{V} {(f_{i}^{k\text{n}})}^x },
\end{equation}
where $C$ controls the overall balance between positive and negative sample weights, and $x$ adjusts the weight differences among items with varying frequency. Here, $f_{v}^{k\text{n}}$ denotes the normalized frequency of item $v$ under behavior $k$, obtained by mapping the frequency $f_v^k$ to the scale of a reference behavior:
\begin{equation} 
f_{v}^{k\text{n}} = f_{v}^{k} \cdot \frac{ I_{v}^{k_{\text{ref}}}}{\sum_{i}^{\mathcal{V}} I_{i}^{k}},
\end{equation}
where $k_{\text{ref}}$ is the reference behavior.

To adaptively enhance the model intensity awareness, we incorporate interaction frequency, a learnable interaction weight matrix, and item representations into the framework:
\begin{equation} 
 f_{v}^{k} = (W_{int} \cdot {q'}_v) \cdot \frac{I_{v}^{k}}{ \sum_{i}^\mathcal{V} {I_{i}^{k}} },
\end{equation}
where $I_v$ is the number of interactions of item $v$, and $W_{int}$ is the learnable weight matrix quantifying the intention intensity.
 
\subsubsection{Joint Multi-Behavior Optimization}
While user behaviors such as view, add, and purchase, are inherently related, modeling them in isolation may disrupt their underlying connections. As such, following the work~\cite{Argyriou2006MultiTaskFL}, we perform multi-task learning to jointly enhance the prediction of multiple behavior types. The definition is as follows:
\begin{equation}
\mathcal{L}_\mathrm{MTL} = \sum_{k=1}^K \lambda_k \mathcal{L}_k(\Theta) + \mu \|\Theta\|^2,
\end{equation}
where $K$ represents the number of behavior types, $\lambda_k$ is the weight of each relationship, and $\mu$ is the regularization term.

Finally, in prediction layer, the interaction probability between users and items is predicted with the following formula:
\begin{equation}
	\hat{y}_{u,v,k} = W_{pre}^{k} ({p'}_u \odot {q'}_v),
\end{equation}
where $W_{pre}^{k}$ represents the learnable weight matrix of the linear predictive transformation weights for behavior type $k$.

\subsection{Time Complexity Analysis}
We present the complete training process of the HiFIRec model in the Algorithm~\ref{alg:HiFIRec}. Given users, items, and the behavior tensor, embeddings for users, items, and behavior edges are initialized (line 1) and updated through L layers of message propagation (lines 2–4). Multi-layer features are fused via attention to obtain refined representations (lines 5–6). For each behavior type, confidence scores are set and behavior-specific losses are computed for all user–item behaviors (lines 7–10). The multi-task loss is aggregated across behaviors (line 11), followed by parameter updates via gradient descent and the prediction of interaction probabilities using the refined embeddings (lines 12–13).

The computational complexity of different components is as follows:

1) Neighborhood aggregation. The propagation for each behavior type requires $d$-dimensions embeddings, and the embedding update involves linear transformations across $K$ behavior types, followed by the propagation of the layer $L$. Therefore, the related complexity is $O(L(|\mathcal{E}|d + |K|d^2))$, where $\mathcal{E}$ is the number of behaviors.

2) Multi behavior feature fusion. The attention weight calculation between users/items and their $K$ associated behaviors requires dot products of $d$-dimensional vectors, resulting in $M \times K$ and $N \times K$ computations. Combined with the $O(d^2)$ matrix operations in fusion, the total complexity becomes $O(MKd^2 + NKd^2)$.

3) Multi-Behavior prediction. It requires calculating negative item weights for each behavior type, involving traversing all $v$ and $K$ behaviors types, with a complexity of $O(K|\mathcal{V}|)$.

HiFIRec employs incremental graph propagation and low-rank multi-behavior fusion, which increase training cost but add no extra inference overhead. After deployment, it supports real-time responses and efficient multi-behavior modeling, making it suitable for lightweight deployment.

\section{Experimental Results}
\label{s:Experimental Results}

To investigate the performance of our proposed HiFIRec, we conduct extensive experiments to answer the following three questions:

\begin{itemize}[left=0pt]
\item RQ1: How does HiFIRec perform compared to the state-of-the-art recommendation methods?

\item RQ2: What is the effect of each individual module on the performance of HiFIRec?

\item RQ3: How do different hyperparameter settings affect the results of the HiFIRec model?
\end{itemize}

\subsection{Datasets and Assessment Metrics}

E-commerce platforms are typical scenarios with high-frequency yet low-intention user behaviors. The associated behavior data is not only vast in volume but also exhibits considerable diversity and temporal dynamics. For example, We validate our method on two real-world datasets \textsc{Beibei} (smart retail platform) and \textsc{Taobao} (mobile e-commerce system), two widely used public datasets in the industry, both fully capture a wide range of user behavior types (e.g., view, add, and purchase). The dataset is constructed with reference to previous research~\cite{chen2021graph}. These datasets provide a realistic testbed for modeling multi-behavior interactions.

\SetAlgoNlRelativeSize{-1}       
\SetNlSty{textbf}{}{}            
\SetNlSkip{1em} 
\SetAlgoInsideSkip{smallskip}    
\SetInd{1.5em}{1em}              

\begin{algorithm}[H]
\caption{\protect\mbox{HiFIRec Training Procedure}}
\label{alg:HiFIRec}
\SetKwInOut{Input}{Input}
\SetKwInOut{Output}{Output}

\Input{\protect\mbox{User set $\mathcal{U}$, item set $\mathcal{V}$, behavior tensor $\mathcal{I}$}}
\Output{Prediction $\hat{y}_{u,v,k}$}

Initialize embeddings $p_u, q_v, w_e$ via Eq.~(1)\;

\For{$l \leftarrow 0$ \KwTo $L-1$}{
    Update $p_u, q_v$ via Eq.~(2)\;
    Update $w_e$ via Eq.~(3)\;
}

Fuse embeddings with attention $\alpha$ via Eqs.~(4--5)\;
Obtain refined $p'_u, q'_v$ via Eq.~(6)\;

\For{$k \leftarrow 1$ \KwTo $K$}{
    \ForEach{user-item pair $(u,v)$}{
        Set confidence $c_v^k$ via Eqs.~(9--11)\;
        Compute loss $\mathcal{L}_k$ via Eq.~(7)\;
    }
}

Aggregate $\mathcal{L}_{\mathrm{MTL}} = \sum_k \mathcal{L}_k$ via Eq.~(12)\;
Update $\Theta$ by gradient descent\;
Compute prediction $\hat{y}_{u,v,k}$ via Eq.~(13)\;

\end{algorithm}

This rich data provides an ideal foundation for investigating differences in user behavior intentions. After data preprocessing, we retain users (or items) with more than $10$ interactions but filter out inactive instances (less than $5$ purchases). Following temporal splitting in~\cite{chen2021graph}, we use the last purchase as test set, the the second-to-last purchase for validation, and remaining interactions for training. Table~\ref{tab:dataset} summarizes dataset statistics. Among them, the ``view'' interaction volume in the Beibei dataset accounts for 71.81\% of the total, and that in the Taobao dataset 77.34\%. The ``view'' behavior accounts for a large proportion of the multi-behavior recommendation data, far higher than other behavior types, and is a typical high-frequency behavior.

We adopted the commonly used evaluation metrics in recommendation tasks, namely HR@K (Hit Rate) and NDCG@K (Normalized Discounted Cumulative Gain). HR@K quantifies the coverage ability of the recommendation list by checking whether the top-K recommended results contain the items that the user has actually interacted with. NDCG@K, on the other hand, evaluates the ranking quality of the recommendation results through a position-weighted gain mechanism. Both metrics have values ranging from 0 to 1, where a higher value indicates better recommendation performance. The values of K are set as K = \{10, 50, 100\}, corresponding to the evaluation of the front page 
\begin{tablehere}
  \centering
  \setlength{\tabcolsep}{3pt}
  \renewcommand{\arraystretch}{1.2}
  \normalsize
  \caption{Datasets statistics}
  \begin{threeparttable}    
    \resizebox{\linewidth}{!}{%
      \begin{tabular}{lcccccc}
        \hline
        \toprule
        Dataset & Users & Items & \#View & \#Add & \#Purchase & Ratio \\
        \midrule
        \hline
        \textsc{Beibei} & 21,716 & 7,977 & 2,412,586 & 642,622 & 304,576 & 71.81\% \\
        \textsc{Taobao} & 48,749 & 39,493 & 1,548,126 & 193,747 & 259,747 & 77.34\% \\
        \bottomrule
        \hline
      \end{tabular}%
    }
    \vspace{0.3em} 
    \caption*{\footnotesize ``Ratio'' represents the proportion of the ``view'' behavior in all behaviors.}
  \end{threeparttable}
  \label{tab:dataset}
\end{tablehere}
recommendation effect, the evaluation of multi-page recommendation lists, and the evaluation of the overall recommendation effect, respectively.

\subsection{Parameter Settings} 
The proposed HiFIRec~\footnote{The code will be released later} is implemented in PyTorch and trained on an NVIDIA 3090Ti GPU with 24GB memory. Optimal parameters are determined using validation data, and performance is evaluated on the test set. We use stochastic gradient descent with adaptive moment estimation (Adam)~\cite{kingma2014adam} for optimization. Hyperparameters are selected via grid search, with batch size chosen from ${8,12,16,20,24}$. The latent factor dimension $d$ is set to 64, and the number of layers is 4. We set the intention intensity range as $C \in [0,1]$, the intention confidence as $x=0.5$, and use ``view'' as the reference behavior $k_{\text{ref}}$.

\subsection{Baselines}
To comprehensively evaluate HiFIRec's effectiveness, we selected five state-of-the-art multi-behavior recommendation methods as baselines. According to whether the baselines use unified behavior modeling or differential behavior modeling, the baselines are divided into two categories.

\noindent\textbf{Unified Behavior Modeling:}

\begin{itemize}[left=0pt]

\item EHCF~\cite{chen2020efficient}: An efficient matrix factorization framework leveraging non-sampling optimization for whole-data learning.
\item GHCF~\cite{chen2021graph}: A heterogeneous graph model with relation-aware GCN capturing multi-hop user-item interactions behaviors.
\end{itemize}

\noindent\textbf{Differential Behavior Modeling:}
\begin{itemize}[left=0pt]
\item S-MBRec~\cite{ijcai2022p285}: A self-supervised learning approach that combines edge dropout with node-level contrastive objectives to learn robust and informative node representations.
\item MBSSL~\cite{xu2023multi}: An advanced self-supervised learning framework incorporating both inter- and intra-behavior contrast and gradient balancing.
\item MBRCC~\cite{lan2024contrastive}: A causal inference model using static causal graphs for exposure bias correction.
\end{itemize}

In our experiments, we implement the public source code and optimize the hyperparameters of all models to achieve the best performance. For the sampling-based method, the negative sample ratio is set to 4\%. For non-sampling methods such as EHCF and GHCF, we assign negative weights of 0.01 for \textsc{Beibei} and 0.1 for \textsc{Taobao}. The loss coefficient $\lambda_k$ is chosen based on the optimal parameters of the GHCF method. In dynamic negative weighting, $x$ is set to 0.5 to balance weight differences between high frequency and low frequency items.

\subsection{\protect\mbox{Overall Performance Comparison (RQ1)}}

\begin{table*}[t]
  \centering
  \setlength{\tabcolsep}{6pt}
  \caption{Performance comparison of recommendation methods on \textsc{Taobao} and \textsc{Beibei} datasets.}
  \begin{threeparttable}
  \resizebox{\textwidth}{!}{%
  \begin{tabular}{clccccccc}
    \hline
    \toprule
    Dataset & Method & Venue & HR@10 & NDCG@10 & HR@50 & NDCG@50 & HR@100 & NDCG@100 \\
    \hline
    \midrule
    \multirow{6}{*}{\textsc{Beibei}} 
    & EHCF~\cite{chen2020efficient} & AAAI'20 & 0.1523 & 0.0817 & 0.3316 & 0.1213 & 0.4312 & 0.1374 \\
    & GHCF~\cite{chen2021graph} & AAAI'21 & 0.1922 & 0.1012 & 0.3794 & 0.1426 & 0.4711 & 0.1575 \\
    & S-MBRec~\cite{ijcai2022p285} & IJCAI'22 & 0.1697 & 0.0872 & 0.3708 & 0.1313 & 0.4604 & 0.1481 \\
    & MBSSL~\cite{xu2023multi} & SIGIR'23 & 0.2229 & 0.1277 & \underline{0.3806} & \underline{0.1626} & 0.4803 & 0.1658 \\
    & MBRCC~\cite{lan2024contrastive} & TOIS'24 & \underline{0.2282} & \underline{0.1323} & 0.3786 & 0.1620 & \underline{0.4805} & \underline{0.1659} \\
    & \textbf{Ours} && \textbf{0.2378} & \textbf{0.1365} & \textbf{0.3925} & \textbf{0.1672} & \textbf{0.4856} & \textbf{0.1679} \\
    \hline
    \cmidrule(lr){2-9}
    & improvement && 4.21\% & 3.17\% & 3.13\% & 2.83\% & 1.06\% & 1.21\% \\
    \hline
    \midrule
    \multirow{6}{*}{\textsc{Taobao}}
    & EHCF~\cite{chen2020efficient} & AAAI'20 & 0.0717 & 0.0403 & 0.1618 & 0.0594 & 0.2211 & 0.0690 \\
    & GHCF~\cite{chen2021graph} & AAAI'21 & 0.0807 & 0.0442 & 0.1892 & 0.0678 & 0.2599 & 0.0792 \\
    & S-MBRec~\cite{ijcai2022p285} & IJCAI'22 & 0.0814 & 0.0446 & 0.1878 & 0.0677 & 0.2605 & 0.0807 \\
    & MBSSL~\cite{xu2023multi} & SIGIR'23 & 0.1027 & \underline{0.0576} & 0.2120 & 0.0813 & 0.3078 & 0.0923 \\
    & MBRCC~\cite{lan2024contrastive} & TOIS'24 & \underline{0.1087} & 0.0492 & \underline{0.2253} & \underline{0.0915} & \underline{0.3155} & \underline{0.0978} \\
    & \textbf{Ours} && \textbf{0.1161} & \textbf{0.0649} & \textbf{0.2548} & \textbf{0.0950} & \textbf{0.3292} & \textbf{0.1071} \\
    \hline
    \cmidrule(lr){2-9}
    & improvement && 6.81\% & 12.67\% & 13.09\% & 3.83\% & 4.34\% & 9.51\% \\
    \hline
    \bottomrule
  \end{tabular}%
  }
  \begin{tablenotes}
    \item Bold numbers indicate the best results, while underlined numbers refer to the suboptimal results.
  \end{tablenotes}
  \end{threeparttable}
  \label{tab:results}
\end{table*}

Table~\ref{tab:results} reports the overall performance of HiFIRec, compared with five state-of-the-art baselines. The results of baselines are collected from the original reported by the authors. Our findings are as follows:

\begin{itemize}[left=0pt]
\item\textbf{Differential behavior modeling paradigm facilitates intention mining.} 
The experimental results show that compared with the unified behavior modeling method (EHCF/GHCF), methods adopting differential behavior modeling (S-MBRec/MBSLSL/MBRCC/HiFIRec) have an average increase of 15.97\% - 23.73\% in the HR@10 metric. This verifies the crucial role of differential behavior modeling in user intentions.

\item\textbf{Noise correction method improves performance.} In noisy signals processing, compared with the method that does not consider noisy signals (S-MBRec), the methods that take into account the noisy signals generated by differences between different behaviors (MBSSL/MBRCC/HiFIRec) have an average increase of 31.35\% - 40.13\% in the HR@10 metric. This targeted design has successfully alleviated model bias caused by behavior differences.

\item\textbf{HiFIRec demonstrates optimal performance in our experiments.} HiFIRec achieves SOTA performance on both datasets with its two stages (network structure and learning strategy). This method introduces a Hierarchical Noise Correction Module and Behavioral Intention Refinement Module in the network structural stage to reduce noisy signals interference and extract high-order features of user intentions. In the learning strategy stage, it introduces a Multi-Behavior Prediction to address the plausible frequent patterns problem by evaluating the intention intensity.

\item\textbf{HiFIRec accurately adapts to data noisy signals in different scenarios.} Cross-dataset experiments reveal an interesting phenomenon: on the Taobao dataset with a higher noisy signals proportion (the ``view'' behavior accounts for 77.34\% compared to 71.81\% in Beibei), the relative improvement of HiFIRec's NDCG@50 reaches 13.09\%, which is higher than 3.13\% in Beibei. This suggests that the denoising effect of the model may be positively correlated with the noisy signals intensity of the data. Further analysis shows that HiFIRec can effectively suppress the noisy signals of high-frequency yet low-intention behaviors and alleviate the effect of noisy signals.
\end{itemize}

Overall, our HiFIRec outperforms existing state-of-the-art methods and shows promising performance in handling high-frequency yet low-intention behaviors. Specifically, differential behavior modeling is beneficial for capturing nuanced user intentions and distinguishing subtle differences across behavior types. Within this paradigm, both the noise correction and intensity-aware non-sampling strategy enhance the model's performance.

\subsection{Ablation Study (RQ2)}

To further gain insight into HiFIRec, we consider five settings for its network structure (i.e., W-NB, P-NB and F-NB) and its learning strategy (i.e., U-NS and I-NS). The details are as follows:

\begin{itemize}[left=0pt]
 	\item \textbf{W-NB (Without Neighborhood Behavior)}: \\This variant relies solely on direct user–item interactions, ignoring higher-order connectivity and contextual signals from neighbors.
 	\item \textbf{P-NB (Partial Neighborhood Behavior)}: \\This variant considers only the direct behavioral interactions between users and items, ignoring the contextual information of the behaviors.
 	\item \textbf{F-NB (Full Neighborhood Behavior)}: \\This variant aggregates neighborhood information, including user–user, item–item, and user–item interactions, to capture user intentions and contextual relationships among nodes.
 	\item \textbf{U-NS (Uniformed Non-Sampling)}: \\This variant assigns fixed weights negative samples uniformly, ignoring the differences in the intensity of intentions implied by different behavior types.
 	\item \textbf{I-NS (Intensity-Aware Non-Sampling)}: \\This variant dynamically assigns weights to negative samples based on the intention intensity of each behavior type.
\end{itemize}

\begin{table*}[t]
  \centering
  \setlength{\tabcolsep}{5pt}
  \caption{Comparison of different embedding propagation modules and optimization strategies.}
  \small   
  \begin{tabular}{c ccccc cccc}
    \toprule
    \hline
    \multirow{2}{*}{Variant} & \multirow{2}{*}{W-NB} & \multirow{2}{*}{P-NB} & 
    \multirow{2}{*}{F-NB} & \multirow{2}{*}{U-NS} & \multirow{2}{*}{I-NS} & 
    \multicolumn{2}{c}{\textsc{Beibei}} & \multicolumn{2}{c}{\textsc{Taobao}} \\
    \cmidrule(lr){7-8} \cmidrule(lr){9-10}
     & & & & & & HR@100 & NDCG@100 & HR@100 & NDCG@100 \\
    \hline
    \midrule
    variant1 & \Large$\bullet$ & \Large$\circ$ & \Large$\circ$ & \Large$\bullet$ & \Large$\circ$ & 0.2751 & 0.0643 & 0.1246 & 0.0316 \\
    variant2 & \Large$\circ$ & \Large$\bullet$ & \Large$\circ$ & \Large$\bullet$ & \Large$\circ$ & 0.4689 & 0.1524 & 0.2613 & 0.0787 \\
    variant3 & \Large$\circ$ & \Large$\circ$ & \Large$\bullet$ & \Large$\bullet$ & \Large$\circ$ & 0.4769 & 0.1620 & 0.2781 & 0.0895 \\
    variant4 & \Large$\bullet$ & \Large$\circ$ & \Large$\circ$ & \Large$\circ$ & \Large$\bullet$ & 0.2947 & 0.0885 & 0.1428 & 0.0575 \\
    variant5 & \Large$\circ$ & \Large$\bullet$ & \Large$\circ$ & \Large$\circ$ & \Large$\bullet$ & 0.4813 & 0.1623 & 0.2674 & 0.0846 \\
    \textbf{Ours} & \Large$\circ$ & \Large$\circ$ & \Large$\bullet$ & \Large$\circ$ & \Large$\bullet$ & \textbf{0.4856} & \textbf{0.1679} & \textbf{0.3292} & \textbf{0.1071} \\
    \hline
    \bottomrule
  \end{tabular}
  \label{tab:ablation}
\end{table*}

Specifically, by pairing a configuration of the network structure with one of the learning strategy, we derive HiFIRec together with five corresponding variants. Table~\ref{tab:ablation} presents the performance of these variants. Our findings are below:
\begin{itemize}[left=0pt]
    \item \textbf{The variants F-NB and I-NS prove the necessity of combining hierarchical noise correction with multi-behavior prediction.} Specifically, on \textsc{Beibei}, the model attains HR@100 of 0.4856 and NDCG@100 of 0.1679, while on \textsc{Taobao}, it reaches HR@100 of 0.3292 and NDCG@100 of 0.1071. This indicates that aggregating neighborhood information, including user-user, item-item, and user-item interactions, enables the model to better capture user intentions. Furthermore, dynamically weighting negative samples by behavior intention intensity more effectively differentiates user intentions.
     \vspace{1em}
    \item \textbf{The variant F-NB proves that our layer-wise neighborhood aggregation design is effective.} This demonstrates its effectiveness in mitigating the effect of noisy signals from high-frequency yet low-intention behaviors. When isolating the effect of full neighborhood aggregation, F-NB consistently outperforms both P-NB and W-NB across datasets. On the \textsc{Beibei} dataset, F-NB achieves an HR@100 of 0.4769 and NDCG@100 of 0.1620, outperforming P-NB (0.4689, 0.1524) and W-NB (0.2751, 0.0643). On \textsc{Taobao}, the improvement is even more pronounced, F-NB achieves an HR@100 of 0.2781 and NDCG@100 of 0.0895, compared to 0.2613 and 0.0787 with P-NB. These results validate our design, where incorporating users, items, and behaviors enhances the model’s ability to capture intentions while reducing behavioral noisy signals.
    \item \textbf{The variant I-NS confirms effective intensity aware non-sampling.} On the \textsc{Beibei} dataset, replacing U-NS with I-NS improves NDCG@100 from 0.0885 to 0.0643, HR@100 also increases from 0.2947 to 0.2751, indicating that intensity-aware weighting helps the model better distinguish the intention intensity reflected by different behavior types. Similarly, on the \textsc{Taobao} dataset, I-NS improves HR@100 from 0.1246 to 0.1428, and NDCG@100 from 0.0316 to 0.0575. These results demonstrate that assigning dynamic sampling weights based on behavior intensity enables the model to more effectively capture user intention, especially in multi-behavior scenarios.
\end{itemize}

The layer-wise neighborhood aggregation and the intensity-aware non-sampling enhance model performance via noisy signals suppression and plausible frequent patterns correction, respectively, enabling more precise user intention modeling.

\subsection{The Effect Of Intention-Sensitive Parameters (RQ3)}
HiFIRec’s frequency correction mainly relies on two intention-sensitive parameters, intention intensity range $C$ and intention confidence $x$. We comprehensively experiment with different combinations of $C$ and $x$ across five ranges of intention intensity values ($C$ = [0 $\sim$ 0.01, 0 $\sim$ 0.05, 0 $\sim$ 0.1, 0 $\sim$ 0.5, 0 $\sim$ 1]) and five intention confidence ($x$ = [0.15, 0.25, 0.5, 0.75, 0.85]). The results on the \textsc{Taobao} and \textsc{Beibei} datasets are shown in Fig.~\ref{Popularity Factor Taobao}. When the distribution difference is too large ($x$ close to 1), the performance drops, likely due to assigning larger weights to pseudo-negative samples. Conversely, when the 

\begin{figurehere}
    \begin{center}
    
      \begin{subfigure}[b]{\linewidth}
        \centering
        \includegraphics[width=\columnwidth]{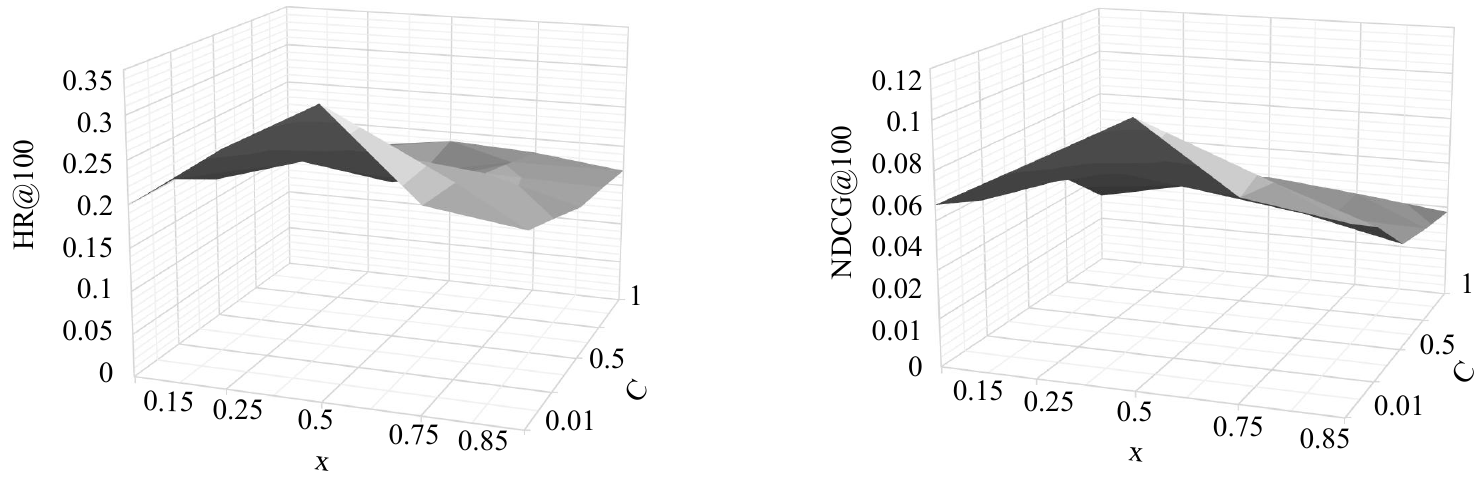}
        \caption{\textsc{Taobao} dataset.}
      \end{subfigure}   
      
      \begin{subfigure}[b]{\linewidth}
        \centering
        \includegraphics[width=\columnwidth]{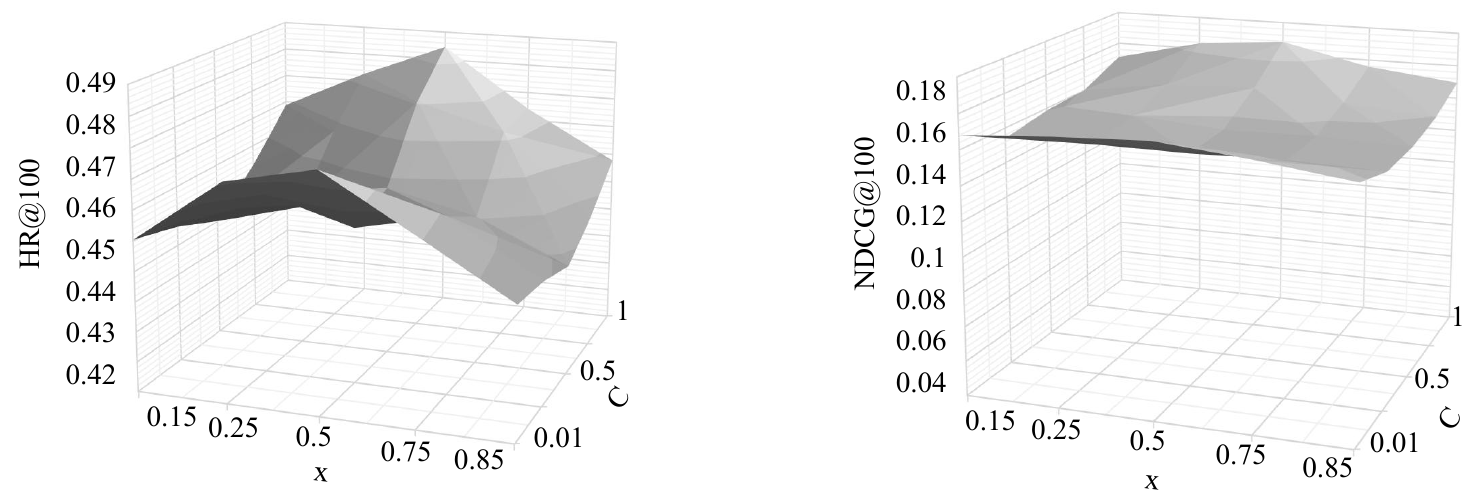}
        \caption{\textsc{Beibei} dataset.}
      \end{subfigure}
      
      \caption{The impact of x and C on the results in the \textsc{Taobao} and \textsc{Beibei} datasets.}
      
      \label{Popularity Factor Taobao}
    \end{center}
\end{figurehere}
difference is too small ($x$ close to 0), the model loses sample distinctiveness. The optimal combination is $C$ = 0 $\sim$ 0.01 and $x$ = 0.5 for \textsc{Taobao}, and $C$ = 0 $\sim$ 1 and $x$ = 0.5 for \textsc{Beibei}.

\subsection{Behavior-Aware Negative Weight Assignment (RQ3)}
\begin{figure*}
  \centering
  \begin{subfigure}[b]{\textwidth}
    \centering
    \includegraphics[width=1\linewidth]{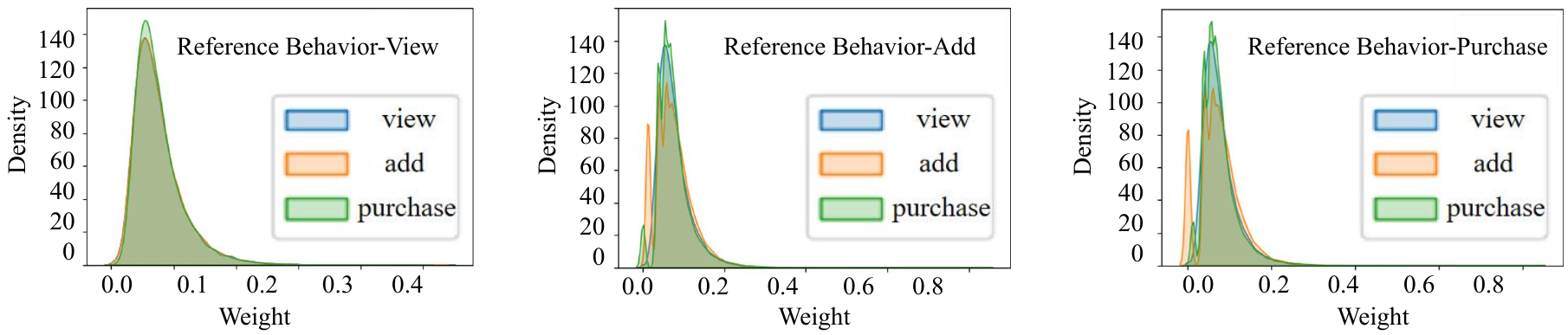}
    \caption{\textsc{Taobao} dataset.}
  \end{subfigure}

  \vspace{1.5em}  

  \begin{subfigure}[b]{\textwidth}
    \centering
    \includegraphics[width=1\linewidth]{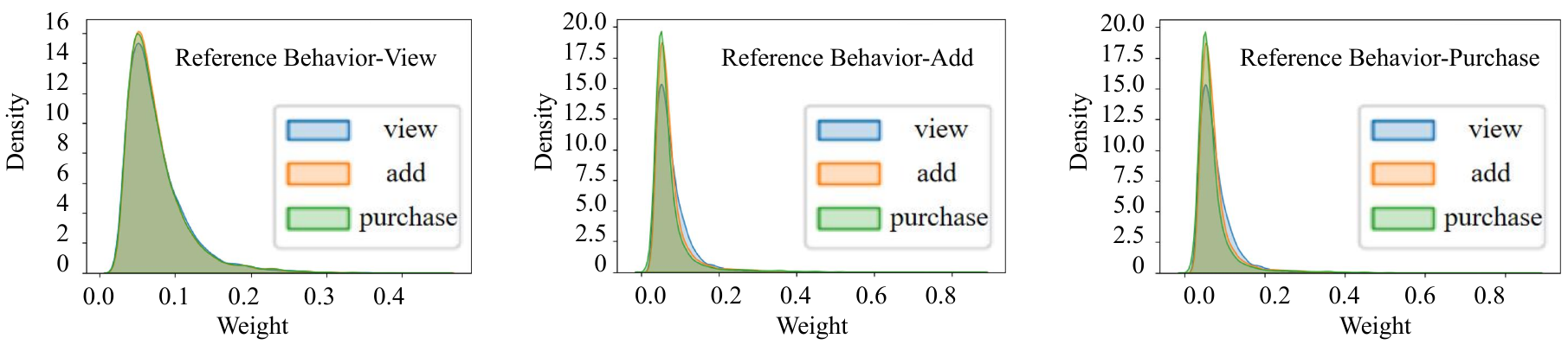}
    \caption{\textsc{Beibei} dataset.}
  \end{subfigure}

  \caption{The distribution of negative weight values for different types of behavior in the \textsc{Taobao} dataset and \textsc{Beibei} dataset.}
  \label{Taobao1andBeibei1}
\end{figure*}

Different behaviors occur at vastly different frequencies and scales, directly using frequency without normalization results in an imbalanced distribution of negative weights. To address this issue, we introduce a reference benchmark behavior $k_{\text{ref}}$ as a unified standard for calculation, and normalize frequency values (Eq.~10) to ensure that the negative weight distribution (Eq.~11) remains smooth and controllable. After determining the optimal values for $C$ and $x$ through preliminary experiments, we conducted independent experiments using ``view'', ``add'', and ``purchase'' as reference behaviors. As shown in Fig.~\ref{Taobao1andBeibei1}, the distribution of negative weights across tasks varies significantly depending on the selected reference behavior. When ``view'' is used as the reference point, the resulting distributions in both data sets are considerably smoother and produce superior performance.

\begin{figure}[H]
  \centering
  \includegraphics[width=0.75\columnwidth]{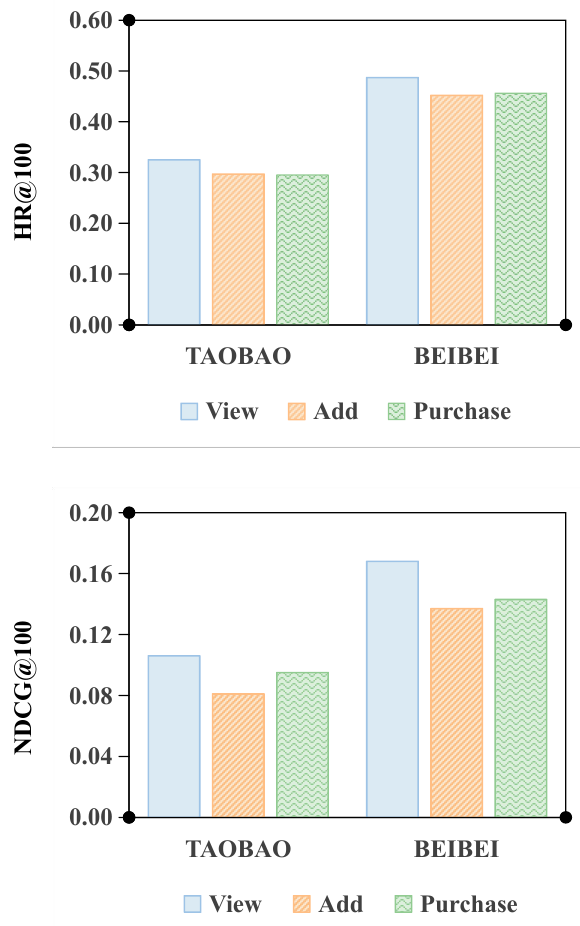}
  \caption{Comparison of negative weight performance of different types of behavior.}
  \label{Popularity Basis}
\end{figure}

The viewing behavior, characterized by broad coverage and high frequency, serves as a stable and representative basis for computing negative weights. This setting not only ensures a more uniform distribution of negative weights but also helps preserve task diversity. In contrast, selecting ``add'' or ``purchase'' as the reference behavior leads to imbalanced weight distributions, which may negatively affect model training. The results in Fig.~\ref{Popularity Basis} further confirm that using ``view'' as the reference point achieves the best performance by maintaining both smooth negative weight distribution and maximum task diversity. In practical shopping scenarios, items that users view but do not purchase typically indicate a lack of purchase intention and should be assigned higher negative weights. In contrast, items that are added but not purchased often reflect only partial or positive intention, and therefore are less reliable as reference behavior for modeling user intentions. Consequently, the ``add'' behavior is not an ideal choice for reference behavior in intention estimation.

\section{Conclusion}
\label{s:Conclusion}

This paper investigates the effect of high-frequency yet low-intention behaviors, which potentially contains noisy signals and plausible frequent patterns. Specifically, we propose HiFIRec that captures behavioral intention from the network structure and learning strategy. We design layer-wise neighborhood aggregation and intensity-aware non-sampling strategies to achieve more effective representations. Extensive experiments demonstrate that our HiFIRec outperforms several state-of-the-art methods, indicating its effectiveness of revising high-frequency yet low-intention behaviors. Future research will explore integrating neighborhood information in multi-modal settings, applying cross-domain approaches, and investigating its scalability and effectiveness across a broader range of recommendation scenarios.

\renewcommand{\bibname}{References}
\bibliographystyle{ws-ijns}   
\bibliography{cite}

\end{multicols}
\end{document}